\documentclass[sigconf]{acmart}

\AtBeginDocument{%
  \providecommand\BibTeX{{%
    \normalfont B\kern-0.5em{\scshape i\kern-0.25em b}\kern-0.8em\TeX}}}

\copyrightyear{2022} 
\acmYear{2022} 
\setcopyright{rightsretained} 
\acmConference[WWW '22 Companion]{Companion Proceedings of the Web Conference 2022}{April 25--29, 2022}{Virtual Event, Lyon, France}
\acmBooktitle{Companion Proceedings of the Web Conference 2022 (WWW '22 Companion), April 25--29, 2022, Virtual Event, Lyon, France}\acmDOI{10.1145/3487553.3524233}
\acmISBN{978-1-4503-9130-6/22/04}

\begin{document}

\title{COCTEAU: an Empathy-Based Tool for Decision-Making}


\author{Andrea Mauri}
\affiliation{%
  \institution{Delft University of Technology}
  \city{Delft}
  \country{The Netherlands}
}
\email{a.mauri@tudelft.nl}

\author{Andrea Tocchetti}
\affiliation{%
  \institution{Politecnico di Milano}
  \city{Milan}
  \country{Italy}
}
\email{andrea.tocchetti@polimi.it}

\author{Lorenzo Corti}
\affiliation{%
  \institution{Delft University of Technology}
  \city{Delft}
  \country{The Netherlands}
}
\email{l.corti@tudelft.nl}

\author{Yen-Chia Hsu}
\affiliation{%
  \institution{Delft University of Technology}
  \city{Delft}
  \country{The Netherlands}
}
\email{Y.Hsu-1@tudelft.nl}

\author{Himanshu Verma}
\affiliation{%
  \institution{Delft University of Technology}
  \city{Delft}
  \country{The Netherlands}
}
\email{h.verma@tudelft.nl}

\author{Marco Brambilla}
\affiliation{%
  \institution{Politecnico di Milano}
  \city{Milan}
  \country{Italy}
}
\email{marco.brambilla@polimi.it}

\renewcommand{\shortauthors}{Mauri and Tocchetti, et al.}

\begin{abstract}
Traditional approaches to data-informed policymaking are often tailored to specific contexts and lack strong citizen involvement and collaboration, which are required to design sustainable policies. We argue the importance of empathy-based methods in the policymaking domain given the successes in diverse settings, such as healthcare and education. In this paper, we introduce COCTEAU (Co-Creating The European Union), a novel framework built on the combination of empathy and gamification to create a tool aimed at strengthening interactions between citizens and policy-makers. We describe our design process and our concrete implementation, which has already undergone preliminary assessments with different stakeholders. Moreover, we briefly report pilot results from the assessment. Finally, we describe the structure and goals of our demonstration regarding the newfound formats and organizational aspects of academic conferences.
\end{abstract}

\ccsdesc[500]{Human-centered computing~Collaborative and social computing}
\ccsdesc[300]{Human-centered computing~Empirical studies in HCI}
\ccsdesc[500]{Information systems~Web applications}

\keywords{Crowdsourcing, empathy, gamification, human-centered computing, decision-making}

\maketitle

\section{Introduction}


Cities are facing sustainability challenges due to rapid population growth. Making cities inclusive, safe, resilient, and sustainable is part of the Sustainable Development Goals\footnote{Link to the UN Sustainable Development Goals \url{https://sdgs.un.org/goals}}. When facing these challenges, policymakers optimize infrastructure, services, or policies using measurable objectives, such as carbon footprint~\cite{Simon2017,gouldson2015accelerating}, or criminality~\cite{toole2011spatiotemporal,mohler2011self}. Typical approaches to policy-oriented data collection include participatory approaches and design methods. Participatory approaches, such as crowdsourced policymaking~\cite{aitamurto2012crowdsourcing} and public consultation~\cite{fishkin2006strategies}), advocate deeper citizen participation and more direct representation. Design methods, such as Community Citizen Science~\cite{hsu2020human}, workshops, and focus group interviews, have the benefits of centering on people's perspectives~\cite{steen2011benefits}.




However, traditional methods can suffer from scalability and hyper-locality problems~\cite{carroll2015reviving}, meaning that the resulting solutions are designed for a specific local context, which can be difficult to generalize. Moreover, these methodologies rely on the experiences of policymakers or designers in integrating conflicting perspectives~\cite{klein1989conflict}. In addition, they may not capture the full information as questionnaires can be incomplete, interviews questions may be loosely formulated, or the participants may not have clear opinions since it can be difficult to imagine the effects or the reasons to implement a policy~\cite{polanyi2009tacit}. Moreover, the individuals engaged through these methods remain unaware of other people’s perspectives, leading to biased choices and difficulty to reach a wider policy acceptance~\cite{clarke2021socio}.




For decades, scholars have studied how forming empathetic relationships between designers and users could result in better products or services~\cite{wrigth2008}. The importance of empathy, defined as ``the intuitive ability to identify with other people’s thoughts and feelings – their motivations, emotional and mental models, values, priorities, preferences, and inner conflicts''~\cite{mcdonagh2006empathic}, has been studied in different domains, such as patients-medic relations~\cite{milcent2021using}, education~\cite{whitford2019empathy,BACHEN201677}, racial bias reduction~\cite{patane2020exploring}, gaming~\cite{gilbert2019assassin,BACHEN201677}, and  design~\cite{segal1997empathic,gasparini2015perspective,yuan2014empathy,mauri2022}. Given the relevancy of empathy-based approaches in these domains, we hypothesize they are applicable in facilitating public conversations on a large city-wide scale among stakeholders involved in the decision and policymaking process.

\textbf{Contribution}: we introduce COCTEAU (Co-Creating The European Union), a web-based gamified application enhancing the interaction between citizens and policymakers. The demo showcases how, in contrast with typical deliberation platforms, COCTEAU can elicit \textbf{empathetic} relations between \textbf{different stakeholders} to collect data about a \textbf{societal issue}. Users engage on the platform in gamified activities where they share thoughts with the community and debate about others' opinions.
\section{COCTEAU}

The main actors engaged in COCTEAU are citizens and policymakers. Citizens engage within the platform to share their thoughts and debate the released scenarios. Policymakers manage the platform by creating, sharing, and maintaining scenarios since they are interested in citizens' thoughts to guide their decision-making process. Researchers in the policymaking field may also be engaged to cover the same role as policymakers.

\subsection{Human-Centered Design Process}
\label{sec:design}

The initial co-design phase of COCTEAU~\cite{tocchetti2020} engaged fifteen researchers working on public policy in a half-day in-person workshop. The researchers were partners of the H2020 TRIGGER (Trends in Global Governance and Europe's Role)\footnote{Link to the TRIGGER homepage \url{https://trigger-project.eu/}} project. The objective was to gather feedback on how the platform's principles would work in a cooperative and interactive environment. We ran this workshop by creating a physical version of the tool (e.g., by concretizing the different steps of our design through pictures, structured documents, notes, etc.) and guiding the participants through opinion-sharing, debating, and convergence processes. The chosen topic was "the impact of artificial intelligence in our daily life". Three examples were provided during the workshop as not every attendant was familiar with the topic. Initially, participants were asked to pick a use case and express their opinions and expectations that motivate their choices. Afterward, people were paired to discuss their thoughts and highlight the strengths and weaknesses of the arguments that possibly led to a change in opinion. Finally, participants were grouped to discuss their thoughts further. Each group converged to one statement, presented at the end of the workshop using text and pictures. The design of our tool is based on the key outcomes summarized hereafter.

\textbf{Engagement \& Gamification}. Designing enjoyable activities for citizens and domain experts is necessary to keep them engaged. One of the most used techniques to enhance and achieve engagement is Gamification. Its goal is to promote people's motivations~\cite{ryandeci2000} towards different activities by using game elements and design techniques. Intrinsic motivation, more concerned with self-improvement, is one of the most effective ways to generate a greater feeling of engagement, eventually leading to a long-lasting commitment. On the other hand, extrinsic motivation is employed to achieve a suitable initial level of engagement and motivates people through separable outcomes (e.g., earning a reward).

\textbf{Community}. COCTEAU aims to develop a community made of proactive people. It is important to focus on collaborative and interactive activities since building a united community is a great way to increase the quality of the content provided by users and their commitment towards shared goals~\cite{hamari2013}. Indeed, relatedness (i.e., the need to feel connected and belonging) has been identified as one of three innate psychological needs~\cite{ryandeci2000}.


\textbf{Empathy}. COCTEAU applies empathy to engage users. We expect citizens to empathize with the thoughts shared by others to expand their opinions on the subject of discussion. This principle influences the design of most activities on the platform by establishing emotions as one of the most relevant elements to engage citizens.

\subsection{Current Design of COCTEAU}

This section describes the design of COCTEAU, derived from the insights gathered in the workshop. Figure~\ref{fig:navigation} shows the process that guides users in sharing their opinion as a combination of an image, a textual description, and an emotion (called \textit{vision}). Inspired by typical co-design methods~\cite{Froukje2005}, the process has the following steps:

\begin{figure*}[h!]
    \centering
    \includegraphics[width=0.9\textwidth]{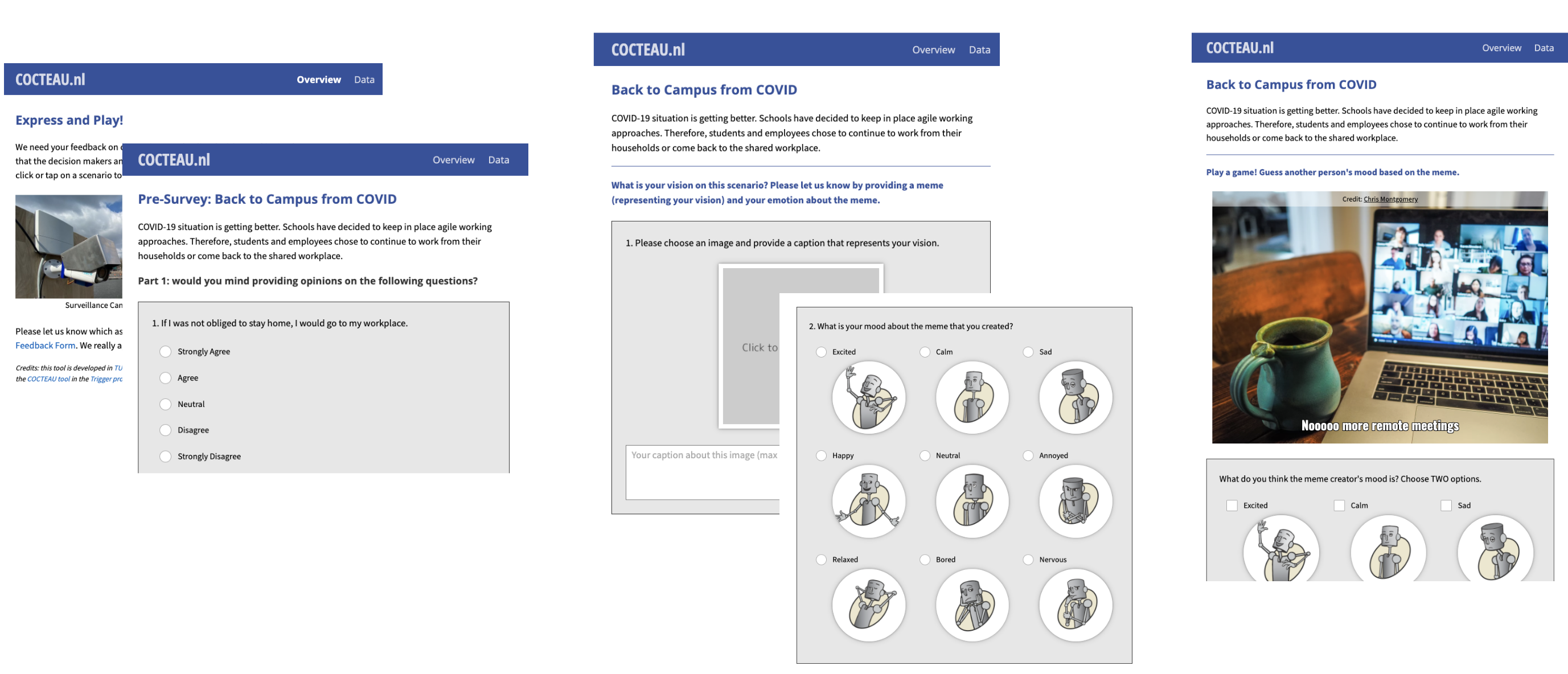}\\
    (a)\hspace{0.6\columnwidth}(b)\hspace{0.6\columnwidth}(c)
    \caption{Three main parts of COCTEAU: sensitivation with a pre-survey (a), content creation (b), and content guessing (c)}
    \label{fig:navigation}
\end{figure*}

\textbf{Sensitization}. As shown in Figure~\ref{fig:navigation}(a), users are presented with a survey. They are asked to assess a series of statements on a 5-point Likert scale from \textit{strongly disagree} to \textit{strongly agree}. The objective of this phase is twofold: first, we collect the starting opinion of the user, and second, we trigger them to reflect on the topic.

\begin{figure}[htpb]
    \centering
    \includegraphics[width=0.86\columnwidth]{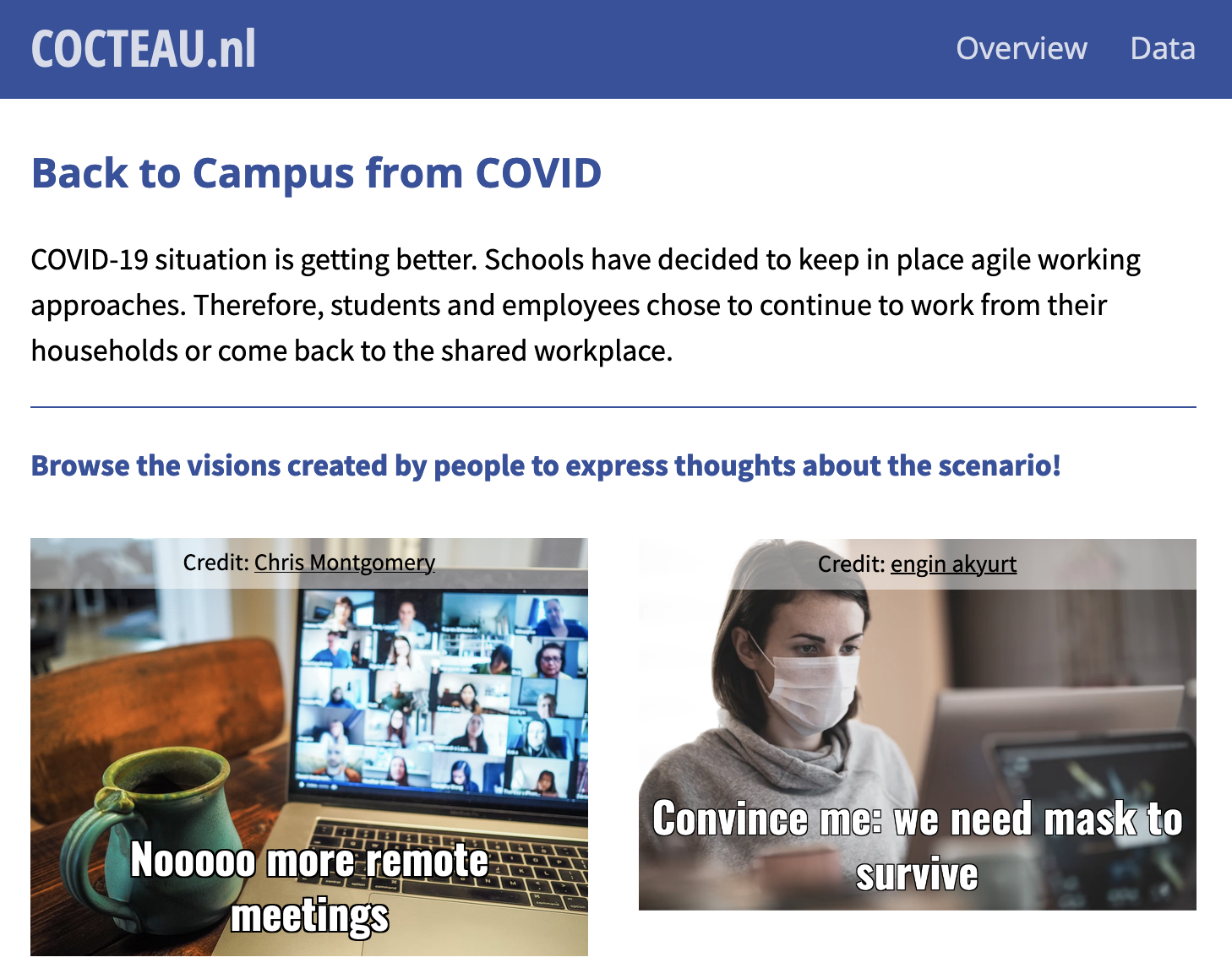}
    \caption{Feed of the visions created by all the users}
    \label{fig:browse}
\end{figure}

\textbf{Content Creation}. During this step, as shown in Figure~\ref{fig:navigation}(b), users create visions by selecting an image, writing a textual description, and picking a mood. We use the Unsplash API\footnote{Link to the Unsplash API -- \url{https://unsplash.com/developers}} for the pictures, allowing keyword-based searches over an image dataset. We use the Pick-A-Mood toolkit (a character-based pictorial scale for reporting and expressing moods) to select the mood, which is a simple yet expressive tool validated in previous studies~\cite{desmet2016mood}. Also, users can browse all visions created by others, as shown in Figure~\ref{fig:browse}.

\textbf{Guessing Game}. In this last step, shown in Figure ~\ref{fig:navigation}(c), users play games where they guess the mood of a vision created by another user. This step allows people to \textit{empathize} with each other since users need to try to understand each other visions. It also engages users through game-play.

\subsection{Implementation}

The platform's implementation is available on GitHub under the MIT license\footnote{Link to our code -- \url{https://github.com/TUD-KInD/COCTEAU-TUD}}. In the documentation of the README file, we explain the techniques and packages that we used to build the platform, which is separated into two parts: the front-end web client and the back-end server. The front-end is implemented using HTML/CSS/Ja-vaScript and the jQuery library\footnote{Link to the jQuery framework -- \url{https://jquery.com}}. The back-end is implemented using Python (with the Flask~\footnote{Link to the Python Flash package -- \url{https://flask.palletsprojects.com/en/2.0.x/}} web framework and the PostgreSQL database\footnote{Link to the PostgreSQL -- \url{https://www.postgresql.org/}}) and deployed using uWSGI\footnote{Link to the uWSGI tool -- \url{https://uwsgi-docs.readthedocs.io/en/latest/}} with an Apache Web server\footnote{Link to the Apache web server -- \url{https://www.apache.org}}. 

\section{Preliminary Evaluation}

We conducted a pilot user study (approved by TU Delft's Human Research Committee) to understand the usability of our tool by engaging students in a design studio.
We applied ethnographic observation to understand how students would use the tool to brainstorm design ideas and concepts.
The objective of the design studio was to conceptualize solutions to urban challenges (health, mobility, sustainability, and tourism) explored throughout the course.
The studio had 25 students, distributed within 5 groups  of 5 students each.
Each group was required to produce a 200 words text describing the challenges, the topics of interest, and the data to support the design process when addressing the challenges.

The studio followed the think-pair-share structure.
Students needed to consider multiple perspectives (such as policy-makers and citizens) by both expressing opinions (think) and discussing the collected perspectives (share).
%
A scenario on COCTEAU was set up with questions related to urban challenges.
At the beginning of the studio, the coach introduced COCTEAU, and all the students were asked to answer the questionnaire on the platform individually.
After that, the coach asked students to check all answers (submitted by others and displayed on COCTEAU) and discuss them with their neighbours.
Then, students were asked to brainstorm ideas with all group members based on the shared answers.
Also, the coach informed students that they could create visions (image with caption) related to the urban challenge the group was interested in exploring.
The coach also showed how they could check others' visions in the tool.
Finally, they were asked to keep working on brainstorming ideas and reaching a consensus on the desired topics.
They were free to choose the combination of tools they preferred to sketch the ideas, such as Miro\footnote{Link to the Miro application -- \url{https://miro.com}}, pen-and-paper, or COCTEAU.

From the observation, students submitted answers using the tool without problems, and no questions were raised regarding usability.
About 50\% of students opened COCTEAU when submitting their answers and used them to support the discussions with their neighbours.
In the group brainstorming stage, the percentage of participants using COCTEAU dropped to about 20\%.
Interestingly, most groups that used pen and paper had COCTEAU open while brainstorming.
This is not the case for the groups that used Miro. 
After the coach informed the students to create visions using COCTEAU, about 30\% of the participants chose to do so.
Noticeably, a group engaged with COCTEAU frequently, and one student in this group specifically played many guessing games without specific instructions from the coach.
%
The student also expressed emotions explicitly when guessing the mood correctly or incorrectly.

There are two major findings from the user study.
First, we noticed the students who used pen and paper (instead of Miro) for brainstorming used our tool more often.
However, it is hard to say if our tool is more effective when the students are using pen and paper, or if it's because with Miro they had to switch between different browser tabs.
Future studies should investigate how COCTEAU could be integrated with existing design and decision-making processes.
%
Second, compared to the ``think'' and ``pair'' phases where the students submitted individual answers and discussed the answers with each other, COCTEAU received less attention during the ``share'' phase.
One possible reason is that the students were concentrated on discussing ideas using Miro or the paper board.
Also, while the content generation phase was thoughtfully designed - as described in Section~\ref{sec:design} - the decision-makers view is still under development.
%
Future studies may need to investigate how COCTEAU can better support the entire brainstorming process.
\section{Demonstration}

The COVID-19 pandemic drastically changed several aspects of our lives, including how we experience academic conferences. Several scientific venues were converted to online-only or hybrid events, providing opportunities to try new ways to organize such events and go beyond the classical conference structure. During the demo, we will ask the attendees to provide opinions on new conference formats (e.g., online v.s. onsite v.s. hybrid, virtual environments such as GatherTown) using our tool. The conference is ideal for our experiment since it involves people from different backgrounds, cultures, and nationalities. There are people from industry and academia of different ranks (e.g., from junior researchers to full professor and department leaders). Each of them has a diverse set of needs and preferences. In parallel, we will showcase the process of setting up the experiment: how to create a scenario and how to define the pre-survey questions. A live version of the tool is available at the following URL\footnote{Link to the demo of our tool -- \url{https://periscope.io.tudelft.nl}}.

\section{Conclusion}

We have presented COCTEAU, a novel web-based tool to engage decision-makers and citizens in gamified activities to collect thoughts about societal issues. The elements and techniques have been co-designed using an in-person workshop with experts in the policy-making field. A preliminary evaluation engaged students from our institution in a series of activities, demonstrating the effectiveness of our platform. The evaluation shed light on the tool's usability, although further investigation is still required. Future works will involve experiments with crowd workers. Decision-makers will also be engaged to design real scenarios to test the tool in a real-world environment and provide relevant outcomes, which will contribute to improving the design and the structure of the platform.

\begin{acks}
This work has been supported by the EU H2020 research and innovation programme, COVID-19 call, under grant agreement No. 101016233 ``PERISCOPE" (\url{https://periscopeproject.eu/}), and the Amsterdam Institute for Advanced Metropolitan Solutions (AMS Institute).
\end{acks}

\bibliographystyle{ACM-Reference-Format}
\bibliography{sample-base}


\end{document}